\documentclass{sig-alternate}
\usepackage{enumitem}
\usepackage{etoolbox}
\usepackage{color,soul}


\begin{document}

\conferenceinfo{AstroHPC'12,} {June 19, 2012, Delft, The Netherlands.} 
\CopyrightYear{2012} 
\crdata{978-1-4503-1338-4/12/06} 
\clubpenalty=10000 
\widowpenalty = 10000

\title{SkuareView: Client-Server Framework for Accessing Extremely Large Radio Astronomy Image Data.}

\def\sharedaffiliation{
\end{tabular}
\begin{tabular}{c}}

\numberofauthors{4}
\author{
\alignauthor 
Vyacheslav V. Kitaeff \\
       \email{slava.kitaeff@icrar.org}
\alignauthor 
Chen Wu,\\
       \email{chen.wu@icrar.org}
\alignauthor 
Andreas Wicenec,\\
       \email{andreas.wicenec@icrar.org}
\and
\alignauthor 
 Andrew D. Cannon\\
       \email{20175501@student.uwa.edu.au}
\alignauthor 
\alignauthor 
Vinsen Kevin,\\
	\email{kevin.vinsen@icrar.org}
\sharedaffiliation
       \affaddr{International Centre for Radio Astronomy Research}\\
       \affaddr{University of Western Australia}\\
       \affaddr{M468, 35 Stirling Hwy, Crawley, WA, Australia}
}

\maketitle

\begin{abstract}
The new wide-field radio telescopes, such as: ASKAP, MWA, and SKA; will produce spectral-imaging data-cubes (SIDC) of unprecedented volume. This requires new approaches to managing and servicing the data to the end-user. We present a new integrated framework based on the JPEG2000/ISO/IEC 15444 standard to address the challenges of working with extremely large SIDC. We also present the developed \textit {j2k} software, that converts and encodes FITS image cubes into JPEG2000 images, paving the way to implementing the presented framework.
\end{abstract}




\section{Introduction}

Currently most radio astronomy data is stored and distributed in the three most commonly used formats, namely: FITS \cite{CFITSIO-IO2011}, Measurement Set (MS hereafter, used mainly in AIPS/CASA environments) \cite{CASA} and HDF5 \cite{Anderson2011, Anderson2010}. FITS and HDF5 are, in general, single self-describing files containing the image data as well as metadata. MS, on the other hand, is a hierarchical structure of directories and files representing multiple frequencies and polarisations, plus the relevant metadata. MS data sets are usually distributed as a single file by using one of the common archiver formats, such as e.g. \textit{tar} \cite{tar}. These formats provide both, portability and access to image data. Normally the spectral-image data-cube (SIDC) is retrieved from an archive and stored on a local computer; and then used as a whole. Alternatively, only part of an image can be extracted as required and downloaded to a local commuter for processing or exploring.

The International Virtual Observatory Alliance (IVOA) has developed the Simple Image Access Protocol (SIAP) standard \cite{IOVA} which defines a protocol for retrieving image data from a variety of astronomical image repositories through an uniform interface. By using SIAP the user can query compliant archives in a standardised manner and retrieve image files in one or multiple formats depending on the archives capabilities (e.g. FITS, PNG or JPEG). The resulting files can then be stored on a local computer or on a virtual network storage device provided through VOSpace, which is another IVOA standard. The approach presented in this paper could be made available through IVOA services, as well, but it also offers extended capabilities which are currently not present in SIAP.

SIDCs from the radio telescopes that are currently under construction -- Australian Square Kilometre Array Pathfinder (ASKAP) \cite{Cornwell2011}, Murchison Widefield Array (MWA) \cite{5164979}, are expected to be in the range of tens of GBs to several TBs. The Square Kilometre Array (SKA) Design Reference Mission: SKA Phase 1 \cite{SPDO2011} defines at least one survey, namely the ``Galaxy Evolution in the Nearby Universe: HI Observations", for which the SKA pipeline will produce hundreds of SIDC of 70-90TB each. If the goal $5.7$ degree$^2$ field of view is achieved with the same resolution, the data-cube size will get close to \(\sim\)1 PB. In one year SKA Phase 1 is going to collect \(\sim\)8 EB of data.

Even taking into account projected advances for HDD/SSD and network technologies, such large SIDC can not be processed, stored or even viewed on local user computers. Therefore such large volumes of data require a new paradigm to generating and servicing the higher level data products to the users or science processing HPC applications. In this paper the authors start to define a new paradigm for working with extremely large radio astronomy images and support it with a suitable framework.

The rest of this paper is structured as follows. In Section 2, we define the necessary functionality and requirements for working with extremely large spectral-imaging data-cubes. In Section 3, we review existing formats commonly used in radio astronomy. In Section 4, we review the JPEG2000/ISO/IEC 15444 standard and discuss its relevance to the defined needs of extremely large images in radio astronomy. In Section 5, we present the \textit{SkuareView} client-server framework for accessing extremely large radio astronomy SIDCs. In Section 6, we describe the \textit{f2j} utility for converting FITS image cubes to J2K files. Finally in Section 7, we summarise and draw the conclusion.

\section{Working with extremely large imaging data}
The extremely large SIDC that will be produced by the next generation of radio telescopes will require new approaches for managing and servicing of the data. In the spectral-imaging domain, a new paradigm is required that satisfies several major requirements to enable the user to work with such large volumes of data, and ease the requirements for data storage at the same time.
\begin{enumerate}[leftmargin=*]
\item The extremely large SIDCs can not be easily downloaded to a local computer. Instead, we believe, that a user should be able to operate with the data in general terms such as: ``survey", ``part of the sky", ``object",``frequency" etc., without a need to specify the file name, its location or having to think about the format in which the data is stored. Files (such as: FITS, HDF5, MS, or other), as means of portability, should only be generated on request to export portions of the data. Such a view will not only simplify user's life, but will also allow to make the internal structure of the data optimised for performance and functionality as required by the intended specific usage of the data. Extremely large SIDCs may not be stored as a single file on a single hard drive. Instead, parts of SIDC might be distributed across multiple hard drives and even network area storage for storage and/or access optimisation.
\item In order to hide the internal structure of the system that is managing extremely large imaging data, a client-server protocol is required. Such a protocol should allow retrieving both metadata and any part of any image from the server.
\item Image data can be effectively compressed. Compression for extremely large images is a ``must", because it will produce significant reduction in the cost of the storage, operations and network bandwidth.
\item Image data, as well as metadata, should be accessible without a need to decompress the whole image. Even in the automated processing of images, i.e. source finding or cross-identification, parts of SIDC need to be randomly accessed. Such random access must be possible without decompressing the whole compressed image. Two levels of compression must be available: 
\begin{itemize}
\item lossless compression - when the decompressed image is an exact reproduction of the original uncompressed image, and;
\item lossy compression - when the decompressed image is only a reproduction of original image with an error.
\end{itemize}
\item Two metrics of errors need to be defined for the compressed radio astronomy images: 
\begin{itemize}
\item statistical characterisation of how the pixels of the decompressed image differ from the original pixels, and;
\item a measure of impact of the compression on the scientific value of the data.
\end{itemize}
The second one is especially important, given that much of the science is done at a very low signal-to-noise ratio (SNR).
\item Full resolution or full fidelity image data in many cases may not be required as a first instance. It should be possible to access multiple lower resolutions and/or multiple reduced fidelities of the master image.
\item The image format and client-server protocol should support a progressive transfer of an image. The user should be able to see the whole image of the selected region queried as soon as the first portion of data is transferred. Every further transferred portion of an image should only further improve the quality of the image. The client-server framework should be intelligent enough not to transfer more data than is necessary for displaying or processing.
\item Whether viewed or processed the image should be transposed and projected into the coordinate system and view of the client. This may not be necessary when a portion of image is exported.
\end{enumerate}

\section{Existing radio astronomy image formats}
Let us now discuss if the existing image formats, commonly used in radio astronomy, can address the requirements given in the previous section to work with extremely large images.

The most commonly used format is FITS, which originated from the need to transport self-describing and portable astronomical datasets. Over the years, FITS gradually evolved to also be used for on-line analysis and data archiving, e.g. writing data to sequential media such as tape devices, which does not represent the current storage mechanism for random access (e.g. SSD, HDD, etc.). In our view, FITS was never designed as an image storage format. Moreover the distinct requirements for data transport, online analysis and data storage are very different. Using a single data format to simultaneously address all three requirements is not optimal. 

FITS data is generally operated on at the ``file" level on a local desktop. As a result, the size of the dataset is limited by the capacity of the underlying filesystem or operating system of the desktop machine. For example, by default, CFITSIO can support a single FITS data-cube up to 2GB, with theoretical possibility to support up to 6TB \cite{CFITSIO-IO2011} on a 64-bit operating system if the library compiled with the corresponding option. For larger files, users have to explicitly split FITS files into smaller tiles and load them one at a time, using the library. In addition, accessing a file larger than the RAM capacity of the machine leads to slow memory-disk swapping. As a result, the FITS application developer must possess particular know-how and understand the nuances \cite{CFITSIO-IO2011} in order to fine tune the performance. 

The Hierarchical Data Format Version 5 (HDF5) provides a generic, abstract data model that enables POSIX-style access to data objects organised hierarchically within a single file. The abstract data model is first mapped onto a linear address space as a standard storage format, which is independent of any storage mediums. The HDF5 library then maps the linear storage format to a specific storage mechanism (i.e. memory, disk, socket). The Low Frequency Array (LOFAR) telescope team developed a radio astronomy data format based on HDF5, which was considered as a ``robust, viable data framework" \cite{Anderson2010} for dealing with intricate complexity and massive volume of the data. While significant progress has been made in the HDF5-based LOFAR ``Radio Sky Image Cubes" specification \cite{Anderson2011}, essential and important issues such as: multi-resolution and image compression have not been addressed. 

Given that large image datasets will become the norm in radio astronomy, we feel that there is an urgent need for a framework to load the burden off users in dealing with large datasets.

\subsection{Image compression}
Image compression using tiled image stored in the FITS binary tables has been recently discussed in \cite{white2012tiled}. An \(n\)-dimensional image is first divided into a rectangular grid of `tiles', each of which is then compressed as a block of data. The resultant compressed code stream is stored in a row of a variable length column in a FITS binary table. The rationale of tiling is to be able to access sub-sections of the image without decompressing the entire image. The convention does not go further to define concrete data paths (e.g. load a tile from the disk to cache, etc.) and data processing pertinent to image compression, decompression processing. This is left for the FITS library (e.g. CFITSIO) to interpret and implement. Storing compressed image tiles in binary table columns appears to be a work-around for image compression that has to rely heavily on the library implementation. In addition, accessing a large image chunk that spans several spatially contiguous tiles may become slower as data columns stored in FITS binary table are generally interleaved \cite{CFITSIO-IO2011}.

Based on this tiled compression convention, CFITSIO currently supports four different compression algorithms - Rice, GZIP, IRAF PLIO, and H-Compress. For obvious reasons no binary compression algorithm is able to produce a significant compression for noisy radio astronomy images. The only proposed image compression algorithm, H-transform, has a potential issue when used with radio astronomy images. Unlike optical astronomy images, which are mostly a sprinkle of bright pixels over an even background, radio astronomy images feature diffuse objects on the background with the residual of imperfect calibration. H-transform does not take advantage of the continuity of pixel values within images, therefore the effectiveness of compression is somewhat reduced.

HDF5 compression is tightly coupled with HDF5 chunks. Each chunk represents a logical subset of the original HDF5 dataset and is compressed/decompressed independently without affecting one another. Unlike the FITS tiled compression convention, HDF5 defines a chunk caching mechanism that allows for an efficient data I/O path from disks to memory. While the chunking mechanism does allow partial compression/decompression on the large dataset, the entire chunk still needs to be copied from the disk into the memory for decompression before the actual read operation can occur. This adds a burden of advanced tuning (e.g. chunk size/shape and cache sizes) \cite{HDF5-chunk2009} which may have to be application or processing task dependant.

\subsection{Metadata access}
FITS supports metadata through the keywords used in the header for HDU, image and table extensions. HDF5 provides a more powerful attribute mechanism to storing metadata for all abstract data model objects such as: dataset, group, links, etc. For large numbers of attributes HDF5 will store them in the heap and index them using a B-tree to support efficient query. The HDF5 library uses a ``dense" storage scheme to deal with metadata with large sizes (\(>\)64KB). However, in order to support efficient metadata access, it is better to use a dedicated performant database rather than embedded indexes within the file format per se.

\subsection{Random access of partial image}
Depending on the method used for linearising image data onto the storage medium, the random access along the spectral/temporal dimension can be sub-optimal. In many radio astronomy surveys, spectral image data-cubes in FITS  are often stored as a ``stack" of image planes. This data organisation scheme is very useful for image viewing on the RA-DEC plane. However, if a user views SIDC on the DEC-Velocity plane for source verification based on a particular Region of Interest's (ROI) spectrum profile, several long seek operations are required to skip most data elements that are out of the small ROI in order to travel across multiple planes. The I/O performance will primarily be bounded by seeking especially when the image gets larger.

Random access along the resolution dimension is currently not supported by common astronomical image formats. Previous efforts in optical astronomy  \cite{fernique2010another} have used the hierarchical spatial decomposition techniques to discretise the resolution hierarchy. Random access on the resolution level is supported by the underlying filesystem. However, this solution completely relies on and is limited by a filesystem's metadata access latency, addressing capacity and storage scalability.

\subsection{Progressive transfer}
In astrometry and photometry, Starck and Murtagh \cite{murtagh2006astronomical} have discussed compression methods that support progressive transfer such as: JPEG, Wavelet, and PMT. The ``progressive" nature lies in the capability to transmit an image at a given resolution or fidelity. While it is desirable that each scale can then be retrieved independently for a given resolution level, two issues are prominent. First, the entire image at this resolution still needs to be fully decompressed before the user can view it.  Second, data that has been transferred at resolution level \(k\) cannot be reused to partially form the resolution level \(k+1\).  To address these issues, the LIVE protocol \cite{murtagh2006astronomical} was developed to allow for \textit{progressive compression} based on region of interest. The image at a given low resolution is first displayed and the user can subsequently ``drill down" within a ROI to improve the resolution of the image. While increasing the resolution, the server only needs to transfer wavelet coefficients relevant to the given ROI to the client, thus achieving progressive transfer. While LIVE has exhibited superior results in both quality and performance, it does not define a file format optimised for large image storage and progressive transfer/decompression. 

\subsection{Client-Server protocol}
The Aladin tool has used the HTTP protocol \cite{fernique2010another} for the communication between server and client for displaying interactive FITS images. The two main issues with using HTTP are: 
\begin{enumerate}
\item The HTTP protocol does not support progressive data transfer and for higher resolutions, a completely new set of diamonds need to be transferred from the server to the client if they are not already in the client cache.
\item It requires extra storage of multiple copies (of different resolution levels) within the same sky areas. 
\end{enumerate}

In the visualisation community, for example, Prohaska et al. \cite{prohaska2004interactive} integrated GridFTP and HDF5 virtual drive as the client-server protocol to transfer a large number of data blocks within a ``remote" HDF5 file located on the server. However, the authors had to customise the HDF5 library, which was not designed for performing I/O over the network.

In summary, we believe that while some useful developments on existing radio astronomy image formats have occurred, there is no one solution which could satisfy all or even most of the requirements for working with the extremely large radio astronomy imaging data in a coherent and self-sufficient way. At the same time, other research and development communities, such as: remote sensing, geographic information systems, medical imaging, have indeed developed interesting techniques which could solve many problems which radio astronomy is about to face with extremely large size imaging data. Due to the limitation of this paper we are unable to provide a comprehensive overview of all relevant techniques, instead we will elaborate on the JPEG2000 or ISO/IEC 15444 standard which we have chosen for evaluation as a promising comprehensive and coherent technology specifically developed to address the issues and challenges of working with large images. 

\section{JPEG2000}
JPEG2000 is an image compression standard and coding system. It was created by the Joint Photographic Experts Group committee in 2000 and published as an ISO/IEC 15444 standard\cite{Taubman02}. The purpose of having a new standard was to address weaknesses in existing image compression standards, and provide new features specifically addressing the issue of working with large images. Considerable effort has been made to ensure that the JPEG2000 codec can be implemented free of royalties. Today, there is a great deal of support of the JPEG2000 standard in both, proprietary and open source software.  JPEG2000 has been successfully used in a number of astronomy applications already, including such science applications as HiRISE (high resolution Mars imaging)\cite{Powell2010} and JHelioviewer (high resolution Sun images) \cite{JHelioviewer}. 

The preceding points led to several key objectives for the new standard. The new standard was expected to allow efficient lossy and lossless compression within a single unified coding framework as well as to provide superior image quality, both objectively and subjectively, at high and low bit rates. It was expected to support additional features such as ROI coding, more flexible file format, at the same time to avoid excessive computational and memory complexity, and excessive need for the bandwidth to view an image.

The main advantage offered by JPEG2000 is the significant flexibility of the codestream. The codestream obtained after compression of an image with JPEG2000 is scalable in nature, meaning that it can be decoded in a number of ways; for instance, by truncating the codestream at any point, a lower resolution or signal-to-noise ratio representation of the image can be attained (scalable compression). By ordering the codestream in various ways, applications can achieve significant performance increases \cite{Taubman02}.

The key main features which make JPEG2000 an attractive alternative to other image formats  currently used in radio astronomy are:
\begin{itemize}[leftmargin=*]
\item Superior compression performance. The previously mentioned problem with the H-transform is solved in JPEG2000 using CDF 5/3 for lossless and CDF 9/7 for lossy compression. Orthogonal Haar wavelet, which is the basis for H-transform, can, in fact, be obtained from CDF transform\cite{CPA:CPA3160450502}.
\item Availability of multi-component transforms including: arbitrary wavelet transforms, arbitrary linear transforms (e.g., KLT, block-wise KLT, etc.) with both reversible and irreversible versions.
\item Multiple resolution representation.
\item Progressive transmission (or recovery) by fidelity or resolution, or both.
\item Lossless and lossy compression in a single compression architecture. Lossless compression is provided by the use of a reversible integer wavelet transform.
\item Random code-stream access and processing, also referred as ROI: JPEG2000 code streams offer several mechanisms to support spatial random access to region of interest access at varying degrees of granularity. This way it is possible to store different parts of the same picture using different quality level.
\item Error resilience -- JPEG2000 is robust to bit errors introduced by communication channels, due to the coding of data in relatively small independent blocks.
\item Flexible file format: The JP2 and JPX file formats allow for handling of both, the frequency-space or colour-space information.
\item Extensive metadata support and handling.
\item Support of volumetric image cubes through JP3D and 3D volumetric compression as part of Part 2 via the extensive multi-component transforms.
\item Interactivity in networked applications as developed in the JPEG2000 Part 9 JPIP protocol. This feature of JPEG2000 deserves a special consideration due to it's utilisation in our proposed framework.
\end{itemize}

\subsection{JPIP}
JPIP is a client/server communication protocol defined in Part 9 of the JPEG2000 suite of standards, officially entitled "Interactivity Tools, APIs and Protocols". It enables a server to transmit only those portions of a JPEG2000 image that are applicable to the immediate client's needs. Using either HTTP or UDP protocols, JPIP enables the client to access contents of the image file including metadata. This capability results in a vast improvement in bandwidth efficiency and speed when performing some very important and valuable image viewing tasks in a client/server environment, while reducing the storage and processing requirements of the client. The larger the images -- and the more constrained the bandwidth between the client and server -- the greater the benefit of JPIP.

JPEG2000 enables the extraction of subsets of an image through three standard compliant image derivation techniques: 1) spatial level, 2) resolution level, and 3) quality level. That is, from a single  compressed image, a user can remotely extract a particular region of the image, a large or small version of the image, or a high or low quality version of the image, or, any combination of those. JPIP can be used to progressively forward images of increasing quality giving the client a view of the image as quickly as possible.

Such features most desirable for the extremely large radio astronomy images, which can hardly be used without examining the metadata and previewing the image at low resolution first, and transferring only the selected parts of the image to a user's computer. This would normally require generating low resolution images, thumbnails and metadata and link them all together in a database. In a system equipped with JPEG2000 and JPIP, however, it is only necessary to store a single file per image; lower resolutions and thumbnails can be extracted directly out of this high-resolution JPEG2000 ``master" image and downloaded. This removes the need to store, manage, and link images of different resolutions in the database, which can be cumbersome. Once the user chooses to view a particular image, only the resolution layer required to view the entire image on the screen is downloaded. The quality layers are downloaded progressively to give the user an image as quickly as possible. When the user zooms into a particular ROI in the image, only that portion of the image is downloaded, and only the minimum resolution is required. Again, the image can be downloaded progressively by the quality layers. The user can continue to zoom into the image until the maximum quality/resolution is reached, and pan across the image; each time downloading only the area of the image being viewed. The user can then scan across different images of the series, maintaining the same ROI and resolution. Again, only the area being viewed is downloaded. The result is a dramatic increase in speed of viewing, and significant increase in the quality and efficiency of the viewing experience.

\subsection{JPIP Stream Type}
The JPIP standard allows three different types of image data to be transmitted between the server and client: 1) full, self-contained JPEG2000 images, 2) tile data, and 3) precinct data \cite{Taubman}.

\textit{Full JPEG2000 Images}. For this data type the server sends to the client complete JPEG2000 images, at the requested resolution. The resolution level is selected to fit in the display window. Because the JPEG2000 images are self-contained, they do not require any additional metadata or headers during transmission; the image is simply sent to the client and the client decodes it.

\textit{Tiles}. Tiles are rectangular spatial regions of images. The image can be encoded to have a single tile or multiple tiles of an arbitrary size. For this data type, the server sends complete tiles to the client, one tile at a time. For tile data, full resolution tiles are always sent. Because tile data is not a self-contained image, additional JPIP messaging headers are attached to convey to the client the contents of the messages.

\textit{Precincts}. JPEG2000 image can be encoded to have one or more precincts per resolution level. Precincts provide an arbitrary spatial subdivision of each resolution level, and are the providers of ROI functionality in JPEG2000. For this data type, the server sends individual precinct data to the client, one precinct at a time. For sub-quality requests, partial precincts can be returned. The server sends only the precincts that intersect the region being viewed by the client at the requested resolution. Additional JPIP messaging headers are attached to the precinct data to convey to the client their contents. This image type is often the most efficient, as it requires the smallest amount of data to be transmitted. In ASKAP and SKA cases, precincts can be defined automatically using catalogues produced by the source finding software as part of the telescope pipeline. Only precincts containing sources are then sent to the client JPIP application at higher resolution; "empty" parts of image can be sent at much lower quality or resolution saving the bandwidth and increasing the speed of fetching and viewing the data. 

\subsection{JPIP Operation and Features}
The client application generates and sents to the server a properly formatted JPIP request containing information about the specific region of the image that the user wishes to view, along with the desired resolution and quality layer data. JPIP server parses the request, calls the JPEG2000 library to extract the relevant image data, generates and cents back to the client a formatted JPIP response. When the response is received by the client, JPIP extracts the image data and the image is recreated using the JPEG2000 decoder. Depending on the settings, the client application can either display the data returned by the JPIP library directly, or -- if JPEG2000 formatted data was specified -- use the JPEG2000 library to decode and display the image.

Tile and precinct databins are the basic elements of a JPEG2000 image used by JPIP. JPEG2000 files can be disassembled into individual finer elements, called \textit{databins}, and then reassembled. Each databin is uniquely identified and has a unique place within a JPEG2000 file. Full or partial databins are transmitted from the server to the client in response to a JPIP request. The JPIP client can decode these databins and generate a partial image for display at any point while still receiving data from the server.

JPIP provides a structure and syntax for caching of databins at the client, and for communication of the contents of this cache between the client and the server. A client may wish to transmit the contents of its cache to the server with every request, or allow the server to maintain its own model of the client cache by maintaining a connection. In either case, the server will reduce the amount of data it is transmitting in response to a JPIP request by eliminating the databins that the client had received in previous transmissions. In this way, JPIP provides a very efficient means of browsing large images in a standard-compliant fashion.

Precinct databins contain all the data from a region of the image at a specific resolution. Precincts are internally divided into packets, which in turn are ordered by quality. In this way the quality of image in each precinct is improved by decoding more and more packets in order.

While databins are being transferred between the server and the client, they usually get split up into smaller chunks, called \textit{messages}. The JPIP server decides the JPIP message size. This flexibility to transmit partial databins enables one to vary the progressive nature of the data being sent to the client. If entire databins are sent in order, the data will be received in a progressive resolution fashion; if messages from different databins at the same resolution level are interlaced, the data will be received by the client with progressive quality. This allows applications to control the user experience depending on the application requirements \cite{Taubman}.

\section{SkuareView}

\textit{SkuareView} is a JPIP based client-server framework aiming to evaluate the potential of JPEG2000 standard to satisfy the image data access and storage requirements for new generation of radio telescopes such as ASKAP/MWA/SKA. Figure~\ref{SkuareView} shows a block diagram of \textit{SkuareView} Server and Client.

\begin{figure*}[!t]
\centering
\includegraphics[scale=1]{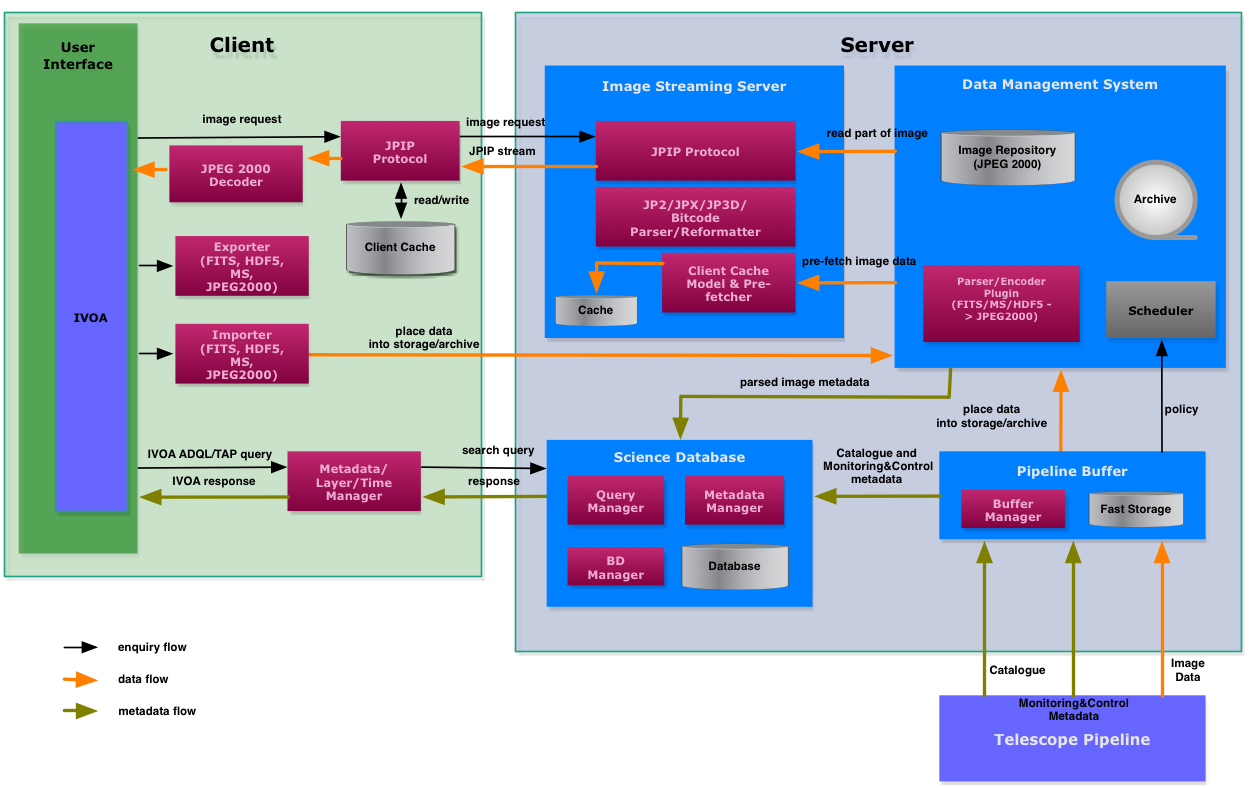}
\caption{SkuareView architecture}
\label{SkuareView}
\end{figure*}

\subsection{Client}
\textit{SkuareView} utilises the IVOA Astronomical Data Query Language (ADQL) and Table Access Protocol for Observation Data Model Core Components \cite{IOVA} protocol to query a \textit{Science Database}. The database contains: a) the catalogues produced by the \textit{Telescope Pipeline} during the surveys, b) metadata extracted and parsed during placement of imaging data into the storage, c) the metadata generated by the \textit{Telescope Pipeline}, d) the metadata generated by the telescope \textit{Control and Monitoring System}, e) information about location of the data files in the storage or archive, and f) possibly some other information related to relevant surveys or science data. IVOA ADQL allows the user to form a query to the database in a uniform platform independent manner.

The \textit{SkuareView Client} can retrieve images from an \textit{Image Repository} in a number of different ways. A typical way would be for the \textit{Client} to 1) query the database, 2) analyse the response and make a decision on what image objects are to be requested, and 3) request an image or precinct(s). 

Examples of a database/image query:\\
-- Retrieve thumbnails of objects from 3$^{\circ}$ region with the centre at ${\alpha}$=3$^{h}$30$^{m}$30$^{s}$ ${\delta}$=10$^{\circ}$10'10" at ${z}$=0.1 to 0.2\\
-- Retrieve time sequenced images given UTC and duration of 5 arcminutes region around of object ${X}$ at resolution 1 arcminute.\\
-- Show a 3D view of an image cube 3$^{\circ}$ region centred at ${\alpha}$=5$^{h}$50$^{m}$50$^{s}$ ${\delta}$=-20$^{\circ}$20'20" at ${z}$=0.1 to 0.5

When a user identifies what imaging data is required the client forms a query to the \textit{Image Repository}. Two options are available to form such a query. 1) IVOA Simple Image Access Protocol (SIAP), and 2) direct JPIP query. In the case when SIAP is used, the query is first translated into a JPIP query. The reason for providing JPIP directly to the client application is that the current edition of SIAP (v1.0) (at the time of writing this paper) does not provide similar extended functionality as JPIP.

The \textit{SkuareView Client} generates a properly formatted JPIP request which contains detailed information about the specific region of the image that the user wishes to view, along with the desired resolution and quality layer data. The request is then sent to the \textit{Image Streaming Server}. The server receives and parses the request, extracts the relevant image data, generates a formatted JPIP response, which is then sent back to \textit{SkuareView Client}.

Other components of the \textit{SkuareView Client} provide the following functionality:

The \textit{JPEG2000 Decoder} and the \textit{JPIP Protocol} represent the client components of JPIP server-client in accordance to JPEG2000 standard. The modules form a request to the server for a part of the image at selected resolution or quality. When a part of the image received from the server, the decoder passes it to client application through User Interface.

The \textit{Client Cache} stores any part of an image received from the server. If a user requests it again, image is retrieved from the cache instead of the server. 

The \textit{Metadata/Layer/Time Manager} component forms a query to the \textit{Science Database} from an IVOA client request and parses the response back to the client application.

The \textit{Importer} and \textit{Exporter} components provide portability functionality of the system. FITS, HDF5, MS or J2K files can be submitted to the \textit{Data Management System} via the \textit{SkuareView Client}. The files are automatically reformatted and encoded to JPEG2000 (if they are in any other format), metadata is extracted, the correspondent records appear in the database. Accordingly, if some of any randomly accessed area needs to be exported from the system, FITS, HDF5, MS or J2K can be produces.

\subsection{Server}
The \textit{Server} includes four main subsystems: \textit{Image Streaming Server}, \textit{Data Management System}, \textit{Science Database}, and \textit{Pipeline Buffer}. Each subsystem can be deployed as centralised or distributed. Generally, blocks on Figure~\ref{SkuareView} reflect functional decomposition rather then deployment. 

The components provide the following functionality:

The \textit{Pipeline Buffer} is not a direct unit of the described framework, however it's an important element to give a generic picture of data movement in the system. The rational for the \textit{Pipeline Buffer} is driven by two factors: 1) the data rate of the pipeline are likely to be too high to make a direct placement of data on the storage or archive, and 2) the data might need to be additionally processed (e.g. reformatted and encoded) before it can be moved to the archive or placed on the ``online" storage or the \textit{Image Repository}.

The \textit{Science Database}, as previously discussed, contains catalogues produces by the \textit{Telescope Pipeline}, extracted during data placement and encoding imaging metadata, metadata generated by the \textit{Telescope Pipeline} and telescope \textit{Control and Monitoring System}, information about the location of the data files in the storage or archive, and some additional information related to relevant surveys or science data. The \textit{Query Manager}, \textit{BD Manager}, and \textit{Metadata Manager} are typical components of a database to provide maintenance, querying and population functionality of a database in the application context.

Since the database contains the catalogue, metadata and location/identifier of images, the possibilities for querying are practically infinite. The database can be used for scientific analysis of the data being supplemented by the images as necessary. The database also holds the history of changes. As the database is explored, it gets populated and improved over time, thus becomes more valuable as it gets used.

The \textit{Data Management System} on its own is a complex system with the main goal of managing the distributed storage of the scientific data, including imaging data. One of the requirements of the \textit{Data Management System} is to provide a fast provisioning of data to the users and HPC applications. In the context of \textit{SkuareView} the two components of the \textit{Data Management System} are essential: the \textit{Image Repository} and the \textit{FITS/MS/HDF5 to JPEG2000 Parser/Encoder Plugin}. The \textit{Image Repository} is a high performance storage system which contains JPEG2000 image files. The \textit{FITS/MS/HDF5 to JPEG2000 Parser/Encoder Plugin} equips the \textit{Data Management System} with the ability to encode and place on the storage various formats of radio astronomy imaging data and access them using of the benefits of JPEG2000 and JPIP. At the same time, the plugin module extracts the relevant metadata and records and passes it to the \textit{Science Database} to be used for querying by users.

The \textit{Image Streaming Server} contains four main components: \textit{JPIP Protocol}, \textit{J2K/Bit-code Parser/Reformatter},  \textit{Client Cache Model \& Pre-fetcher}, and \textit{Cache}. The first two components are defined by  JPEG2000 Part 9 to provide the server side of JPIP streaming. JPEG2000 also, optionally, specifies the \textit{Client Cache Model} on the server side. The \textit{Client Cache Model} can be very effectively used in conjunction with the policies of hierarchical data placement of the \textit{Data Management System}. Based on the client's data usage pattern, it would be possible to achieve pre-fetching of image data from a slower type of media onto a faster \textit{Cache}, if it is predicted that this data is likely to be requested soon by the user or application.

\section{Software to convert FITS SIDC into JPEG2000 images}
As an part of the evaluation and prototype development, the \textit{f2j} software was developed to convert FITS image cubes to JPEG2000 files. The utility is written in C/C++ utilising the open source OpenJPEG codec for JPEG2000 compression and NASA's CFITSIO library for FITS files.

The primary design goals of the \textit{f2j} include efficiency, extensibility and utility as part of the \textit{Data Management System} to support JPIP client-server architecture for viewing extra large radioastronomy SIDC. While the \textit{f2j} can be used as a standalone utility, most of the developed code is reusable for the \textit{FITS/MS/HDF5 to JPEG2000 Parser/Encoder Plugin} module of the \textit{Data Management System}, the interface of which is to be defined separately. 

\textit{f2j} encodes FITS files as JPEG2000 images with a single component consisting of (greyscale) pixel intensities stored as 16 or 32 bit unsigned integers. \textit{f2j} allows optionally to output each plane of FITS cube into a separate JPEG2000 image file or encode all planes into one volumetric JPEG2000 image using multi-component transfer. The JPEG2000 image can be output with different scaling: linear, square root, logarithmic or power.

The encoding has a full range of parameters to control compression type (reversible or irreversible), compression ratio or quality, resolution levels, tile sizes, etc. A residual image -- a difference between the original bitmap and the bitmap after reconstructed compressed image -- can be output and its PSNR value is calculated as a metric of added during compression noise. Non-gaussianity tests can be performed on the residual image as a metric of introduced distortion.

\textit{f2j} software has been optimised and parallelised for multi-core processor architecture (including Intel MIC) using the Intel Thread Building Blocks library.

Future development of \textit{j2k} will add MS images to JPEG2000 conversion, conversion of a whole image cube into a volumetric JP3D, and an option to use Kakadu JPEG2000 library instead of OpenJPEG. 

The source code for \textit{f2j} can be downloaded from GitHub\footnote {https://github.com/ICRAR/SkuareView}.

\section{Conclusion}
We have discussed the challenges of working with extremely large imaging data expected from the radio telescopes of new generation (ASKAP/MWA/SKA). We have defined the necessary functionality and requirements to a system which would enable overcoming the difficulties caused by extremely large size of the images.

We have reviewed the existing image formats used in radio astronomy and demonstrated their limited ability to provide the desired functionality. 

We have reviewed JPEG2000/ISO/IEC 15444 standard and demonstrated the relevance and efficiency of JPEG2000 solutions to extremely large images in radio astronomy. 

We have developed and presented the JPEG2000 based \textit{SkuareVeiw} client-server framework for accessing extremely large radio astronomy images which can provide an effective and efficient solution to working with extremely large imaging data which is expected from new radio telescopes, such as ASKAP/MWA/SKA. Amongst many features and benefits of the JPEG2000-based \textit{SkuareView} framework, we would like to emphasise the following:
\begin{itemize}[leftmargin=*]
\item Significantly reduced storage requirements, and therefore reduced initial and operational cost of the system.
\item Low local disk requirements on the client.
\item  Very high flexibility of structuring the image data (files, tiles, precincts) for distributed data placement.
\item  Ability to view a whole image or part of an image data-cube without a need to decompress the data first.
\item  Ability to handle and access a complex metadata without retrieving the image.
\item  Random access to multiple regions of interests simultaneously without a need to produce file cutouts.
\item  Viewing the images at multiple resolutions and quality (fidelity) as required by the specific science or processing needs.
\item  Low requirements to bandwidth due to the fast progressive transfer at the optimal resolution of only needed data.
\item  Integration with the science and technical meta data in the database that provides powerful querying capability.
\item  Integration with the Data Management System that provides high level of integrity of distributed data.
\item  Portability of various image data formats to the system and ability to flexibly export data from the system based on the area of interest rather than on original container (file or directories and files) of data.
\end{itemize}

We have also developed and presented \textit{j2k} software for converting FITS spectral-image data-cubes to J2K files.

\bibliographystyle{abbrv}
\bibliography{astro04-kitaeff}

\balancecolumns
\end{document}